**Notice:** This manuscript has been authored by UT-Battelle, LLC, under Contract No. DE-AC05OOOR22725 with the U.S. Department of Energy. The United States Government retains and the publisher, by accepting the article for publication, acknowledges that the United States Government retains a non-exclusive, paid-up, irrevocable, world-wide license to publish or reproduce the published form of this manuscript, or allow others to do so, for the United States Government purposes. The Department of Energy will provide public access to these results of federally sponsored research in accordance with the DOE Public Access Plan (http://energy.gov/downloads/doe-public-access-plan).

# Learning and predicting photonic responses of plasmonic nanoparticle assemblies via dual variational autoencoders

Muammer Y. Yaman[1], Sergei V. Kalinin[2], Kathryn N. Guye[1], David Ginger[1,3]*, Maxim Ziatdinov[4,5]*

1. Department of Chemistry, University of Washington, Seattle, WA, USA

2. Department of Materials Science and Engineering, University of Tennessee, Knoxville, TN, USA

3. Physical Sciences Division, Physical and Computational Sciences Directorate, Pacific Northwest National Laboratory, Richland, WA, USA

4. Center for Nanophase Materials Sciences, Oak Ridge National Laboratory, Oak Ridge, TN, USA

5. Computational Sciences and Engineering Division, Oak Ridge National Laboratory, Oak Ridge, TN, USA

* = Corresponding Authors

**Abstract:**
We demonstrate the application of machine learning for rapid and accurate extraction of plasmonic particles cluster geometries from hyperspectral image data via a dual variational autoencoder (dual-VAE). In this approach, the information is shared between the latent spaces of two VAEs acting on the particle shape data and spectral data, respectively, but enforcing a common encoding on the shape-spectra pairs. We show that this approach can establish the relationship between the geometric characteristics of nanoparticles and their far-field photonic responses, demonstrating that we can use hyperspectral darkfield microscopy to accurately predict the geometry (number of particles, arrangement) of a multiparticle assemblies below the diffraction limit in an automated fashion with high fidelity (for monomers (0.96), dimers (0.86), and trimers (0.58). This approach of building structure-property relationships via shared encoding is universal and should have applications to a broader range of materials science and physics problems in imaging of both molecular and nanomaterial systems.

**Introduction**

Assemblies of plasmonic nanoparticles lead to emergent photonic behaviors with applications in fields as diverse as energy harvesting[1–6], medical diagnosis[7–9], sensing[10–14], and catalysis.[15,16] Directed assembly of plasmonic nanoparticles has proven effective in synthesizing hierarchical materials with such desired functionalities.[17–20] Generally, minute changes in the assembly state of the nanoparticles may result in significant changes of the far-field optical response accessible to macroscopic characterization.[21–23] Hence, it is of interest to study whether the optical changes detected via far-field measurements can be used to probe and understand the assembly process, and whether we can predict the optical response of a material system with a known geometry without a complete first principles calculation.

While electron microscopy remains the standard method of probing the assembly process, and of characterizing the resulting structure,[24] the assembly-dependent optical properties of plasmonic nanoparticles also allow the assembly process to be probed directly, in solution,[25] and in near-real time,[26] without the challenges or perturbations that may arise from electron beam irradiation.[27,28] Indeed, over the years interparticle plasmonic coupling has been used to probe the evolution of biological structure, particularly with plasmon rulers.[29,30] Going beyond simple dimers, 3D plasmonic rulers have also been proposed in order to probe 3D structure and even demonstrated with lithographically fabricated nanostructures.[31]

More recently, machine learning (ML) methods have emerged as powerful tools to predict and encode complex structure-function relationships, such as those that arise through the coupling of multiple plasmonic particles.[32–34] Conversely, it should also be possible to use ML methods to predict/reconstruct the sub-wavelength structural information encoded in a complex plasmonic scattering spectrum.[32,35,36] At the same time, applications of ML methods to experimental datasets are limited by the relative scarcity of experimentally accessible data, which is often comparable or smaller than the intrinsic data dimensionality. This sparsity is a ubiquitous challenge facing scientific researchers, limiting the applicability of the classic "big data" supervised learning methods, and thus necessitating the development of novel approaches for building structure-property relationships with a small (compared to intrinsic data dimensionality) number of examples.

Here we develop and demonstrate an approach for connecting the structure-property relationships between far-field optical responses and local nanoparticle cluster geometries in systems of plasmonic gold nanoparticles. We base this approach on dual variational autoencoders for the imaging and spectral data with a shared latent encoding. This approach allows the prediction of the optical properties of a cluster from a known geometry, and also the prediction of the geometry of a cluster from its measured optical response. We apply these methods both to reconstruct and classify the predicted geometries of plasmonic nanoparticle clusters based on their scattering spectra as well as to predict the scattering spectra of resulting clusters.

**Result and Discussion**

As a model system, we examine the plasmonic response of self-assembled, citrate-capped gold nanoparticle clusters, comprising primarily spherical particles with an average diameter of 100 nm. We assemble these gold nanoparticle clusters into different geometries using a salt-induced aggregation approach,[37] as described in detail in the Methods section. These assembled gold clusters were precipitated onto an indium-doped tin oxide (ITO) substrate. The samples were analyzed under both scanning electron microscopy (SEM) and hyperspectral darkfield microscopy. Due to their different geometries, we expect the nanoparticle clusters to generate unique scattering spectra that depends on the number of particles and interparticle spacings, with some consideration to particle shape variation.[38–40]

Fig. 1a depicts our basic approach. In order to relate the geometries of the clusters with their associated plasmonic responses, we collect both high-resolution SEM images, as well as spatially correlated hyperspectral images. By correlating the structures from SEM with spectra from hyperspectral images using the variational autoencoder, we aim to encode the structure-property relationships in the neural networks weights, thereby enabling a prediction of multimer structures based solely on scattering spectral data, and vice versa. To obtain the spectrum of each nanoparticle cluster, we start from identifying the positions of these clusters from hyperspectral images by applying a standard thresholding method, as described in SI section, and then extract the corresponding spectra data at each known position.

Figures 1b and 1c show an example of scattering spectra and SEM images of the nanoparticle clusters respectively. The full dataset, methods, and algorithms are available for download via https://github.com/ziatdinovmax/dualVAE. We note that the stochasticity of the nanoparticle aggregation process leads to clusters with varying shape, size and orientation, and therefore with various optical responses (varying darkfield scattering spectra). Fig. 1b, sub image 2, shows a scattering spectrum characteristic of single gold nanosphere which exhibits a single, Lorentzian plasmon scattering peak at 568 nm, as expected for a 100 nm particle on a glass substrate in air.[41] Indeed, the corresponding SEM image (Fig. 1b, sub image 2) confirms the spectrum came from a single sphere. In contrast, Fig. 1b, sub-image 6, shows a scattering spectrum exhibiting a transverse plasmon resonance peak at 560 nm and a longitudinal plasmon resonance peak at 850 nm. This dual peak is consistent with the expected scattering spectrum of a dimer of 100 nm gold nanospheres (see Fig. S1).[42] The SEM in Fig. 1b/c sub-image 2 likewise confirms this cluster geometry assignment. For other dimers observed, the exact position of the longitudinal peak depends on the interparticle spacing of the dimer as expected (Fig. 1b/c sub-images 8 and 26).

For trimers and larger clusters, the spectra become increasingly complex (c.f. sub images 1, 4, and 30), again this increase in complexity is expected, since the size, shape, number and orientation of the gold nanoparticles all affect the resulting optical response.[43]

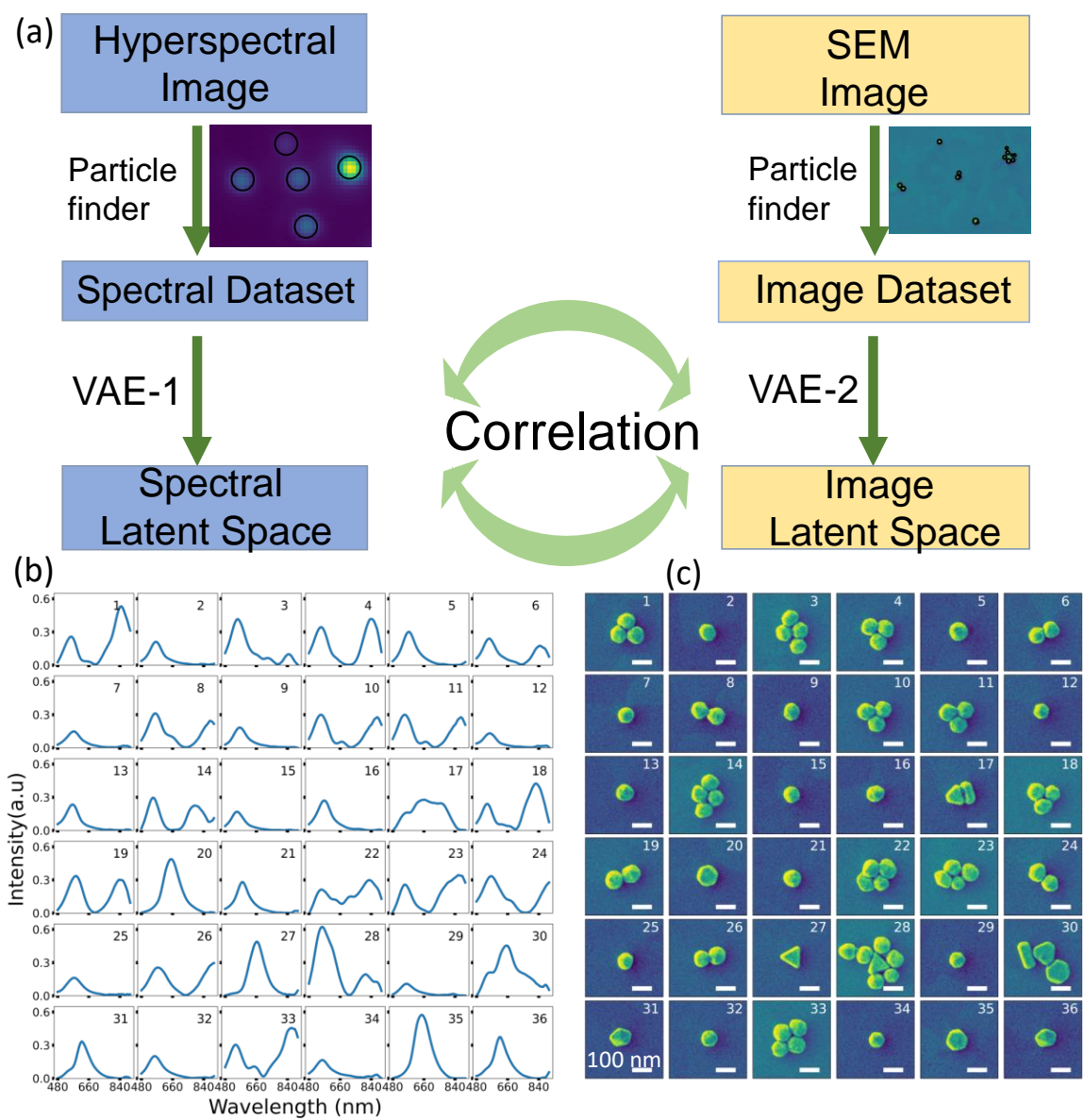

**Figure 1.** (a) Schematic representation of the dataset identification and correlation workflow. (b) Representative spectral data set and (c) their corresponding SEM images. Sub image numbers are shown right top of each image.

In our first analysis, we tried to independently learn the latent representations of the spectral and structural dataset, and subsequently identify the correlation between them. For the spectral data, we explored the applicability of the VAE with different invariances as we have applied previously for scanning probe microscopy[44] and electron energy loss spectroscopy datasets.[45] Figure S1a/c shows the performance of the VAE in this case. However, we see the latent variables ($z_1$, $z_2$) for the spectral data cannot be differentiated from each other.

Due to the stochastic deposition of clusters, the image dataset necessarily comprises images with significant orientation disorder and possible offsets. In these cases, the standard ML approaches based on direct image analyses using linear decomposition methods, such as principal component analysis and standard deep convolutional neural networks, are known to fail.[46] To address this issue, we have previously implemented a family of the rotation-, shift-, and scale-invariant variational autoencoders for unsupervised-,[44,45,47–54] semi supervised-[55] and joint-learning[56] of disentangled latent representations. These VAE-based approaches allowing for the intrinsic physical invariances in the datasets, or imaging process, have successfully extracted order parameters in disordered systems including chemical transformations on the atomic level [46] and organization in protein nanoparticles,[47] as well as disentangled the domain switching mechanisms in ferroelectric materials,[48] suggesting this approach is fairly universal and could be fruitfully applied to the current problem.

For the structural data, we applied a variational autoencoders (VAE) with translational and rotational invariances to disentangle the latent representations. Figure S2 b/d shows the performance of VAE with this structural dataset. One can see the importance of rotationally and translationally invariant VAE from the latent space distribution: they are differentiated based on the number of the particles in cluster ($z_{1,\text{structure}}$ first variable) and cluster shape ($z_{2,\text{structure}}$ second variable), respectively. However, the VAE model cannot separate the spectral dataset well, in Figure S2 a/c. The first variable ($z_{1,\text{spectra}}$) of the VAE model has no change along horizontal axis, whereas the second variable ($z_{2,\text{spectra}}$) shows only a slight change in the peak at 840 nm. Furthermore, when we tried to correlate the latent spaces – using the spectral data to predict the structural data, or vice versa, we were unsuccessful and we found (see Figure S2 e/f) that their latent representation showed no trend, meaning that these two datasets have no correlation.[57]

Because this straightforward approach of trying to correlate the *separate* latent spaces of the spectral and image spaces produces poor results, we next turn to develop the dual-VAE approach for the establishing the structure-property relationships. In a dual-VAE, there is a "communication channel" between latent spaces of two VAE models. Since the data is available in the form of image-spectrum pairs, this correlation allows for several opportunities to control the encoding process. As one such approach, we enforce similarity between structure and spectra latent spaces.

The implementation of dual-VAE is outlined in Algorithm 1. The dual-VAE model consists of two encoders and two decoders. At each training step, the first encoder (*encoder-1*) takes the image data to produce a latent vector, which is then passed through one of the decoders (*decoder-1*) to get a reconstruction of original data. The second encoder (*encoder-2*) takes spectral data to obtain its latent representation which is transformed via a learnable linear transformation before being reconstructed by the second decoder (*decoder-2*). The total loss function is given by:

$$\mathcal{L} = \mathcal{L}_{\text{RE}} + \beta \mathcal{L}_{\text{KLD}} + \gamma L^1$$

where $\mathcal{L}_{\text{RE}}$ is a sum of reconstruction errors for decoded image and spectral data, $\mathcal{L}_{\text{KLD}}$ is a sum of Kullback-Leibler divergence terms[58] between standard normal distribution and encoded distributions of image and spectral data, and $L^1$ is an L1 score between the latent representation of the *encoder-1* and transformed latent representation of the *encoder-2*. The coefficients $\beta$ and $\gamma$ are constant scale factors. The presence of a learnable linear transformation between the two latent representations and the loss term reflecting the 'closeness' of the two representations at each training step allows the alignment of the image and spectral latent spaces for establishing structure-property relationships in the system. We used a standard Adam optimizer for adjusting weights of all the encoders and decoders simultaneously in the end-to-end fashion.[59] The approach outlined in Algorithm 1 works both with standard VAEs and invariant VAEs.

**Algorithm 1: Training of dual-VAE (single step)**

Inputs: Two experimental datasets, $X_1$ and $X_2$.

Pass $x_1$ through *encoder-1* to get $\mu_1$ and $\sigma_1$ parameters of variational distribution
Sample latent vector, $z_1 \sim \mathcal{N}(\mu_1, \sigma_1^2)$
Compute KL divergence, $D_1$, between encoded and prior distributions
Pass $z_1$ through *decoder-1* to obtain $x_1'$
Compute reconstruction loss, $RE_1$, between $x_1'$ and $x_1$

Pass $x_2$ through *encoder-2* to get $\mu_2$ and $\sigma_2$ parameters of variational distribution
Apply a learnable linear transformation, $A$, such that $\mu_2' = A\mu_2$.
Sample latent vector, $z_2' \sim \mathcal{N}(\mu_2', \sigma_2^2)$
Compute $L^1$ score between two latent vectors, $z_1$ and $z_2'$
Compute KL divergence, $D_2$, between encoded and prior distributions
Pass $z_2'$ through *decoder-2* to obtain $x_2'$
Compute reconstruction loss, $RE_2$, between $x_2'$ and $x_2$

Compute total loss, $\mathcal{L} = (RE_1 + RE_2) + \beta(D_1 + D_2) + \gamma L^1$
Backpropagate loss and adjust weights in both VAE models

At the prediction stage, the alignment of latent spaces allows predicting spectra from images and vice versa. In this case, the images/spectra are encoded via encoder-1/encoder-2 and decoded into spectra/image via decoder-2/decoder-1 (Figure 2).

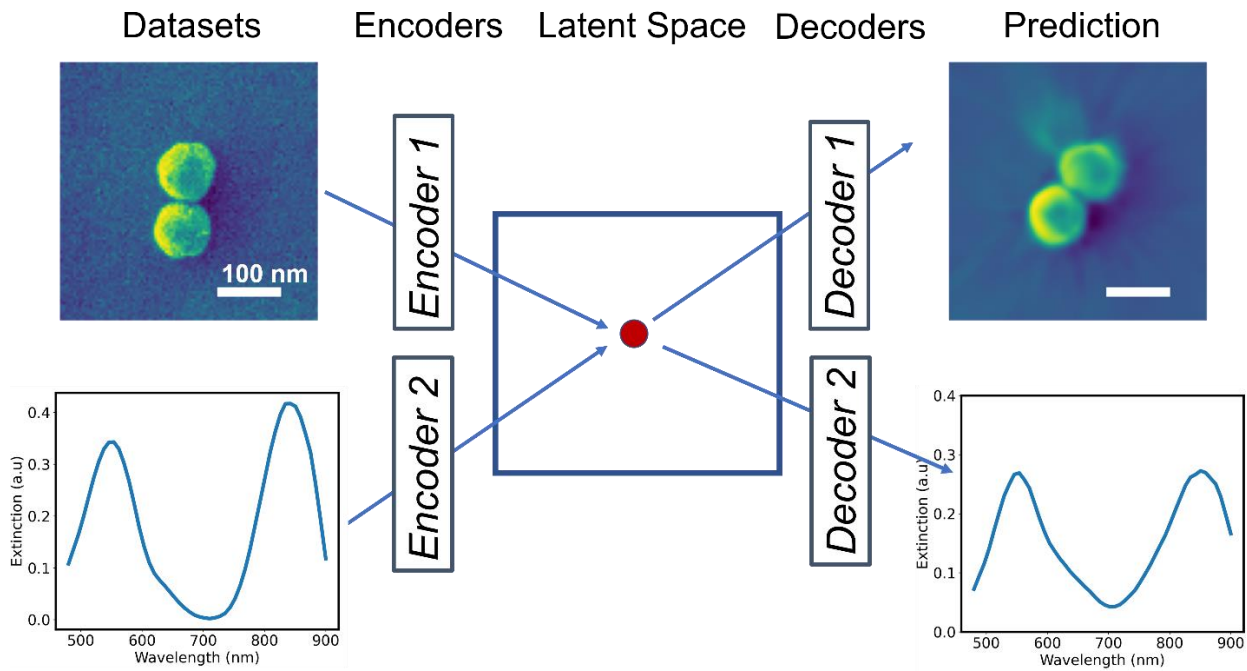

**Figure 2.** Schematic representation of the dual variational autoencoder use for prediction. The left side shows the inputs and right side shows predictions via the shared latent space. The scale bars is 100 nm. Note that when using rotationally invariant VAE for spatial data, all particles of the same shape (e.g., dimers) are aligned in the same direction in the decoded data, independently of the input orientation.

As discussed above, we created an experimental dataset library using the structure (from SEM) and spectral properties (from hyperspectral imagery) information of the gold nanoparticle clusters. This library is constructed of 898 clusters (526 monomers, 197 dimers, 115 trimers, 60 larger clusters ($n \geq 4$ where $n$ refers to the number of particles in a cluster)). We used randomly chosen 247 clusters (138 monomers. 58 dimers, 26 trimers, and 25 larger clusters) to train the dual-VAE model. Figure 3 shows the results generated from the dual-VAE. Here, the structural and spectral data of the clusters are displayed as circle- and star-shape points respectively. Each data point is post-colored with red, green, blue and purple, representing the structure information of monomers, dimers, trimers, and larger clusters respectively. In contrast to the simple VAE approach, (Figure S2), we can see that clusters consisting of different numbers of nanoparticles are well separated in both the structural and spectral latent spaces in the shared latent space, Figure 3a.

Figure 3b shows a plot of the first variable of the spectral latent space as a function of the first variable of the structural latent spaces. Importantly, this plot of $z_{1,spectra}$ vs. $z_{1,structure}$ shows a linear correlation, meaning that these variables are related. Since $z_{1,structure}$ encodes the number of particles in a cluster, the correlation of $z_{1,spectra}$ with $z_{1,structure}$ in Fig. 3b implies that the dual-VAE with a shared latent space has likely learned that the spectra depend on the number of particles in the cluster, as our physical understanding of the system tells us is the case (indeed total scattering intensity should scale with the number of particles in the cluster). [60]

We test this assumption by decoding the latent space into spectral (Figure 3c) and structural (Figure 3d) manifolds. We note that: (1) in Figure 3e, the spectral intensity increases along the horizontal direction (from right to left) which, as expected, matches well with the increase in the number of nanoparticles comprising the clusters in Figure 3f. (2) in Figure 3e, the second peak around 800 nm become more obvious along the vertical direction (from bottom to top), which also corresponds well with the cluster shape change shown in Figure 3f. Notably then, the dual-VAE is able to distinguish not only the particle number in a cluster, but also different cluster shape in both spectral and structural datasets.

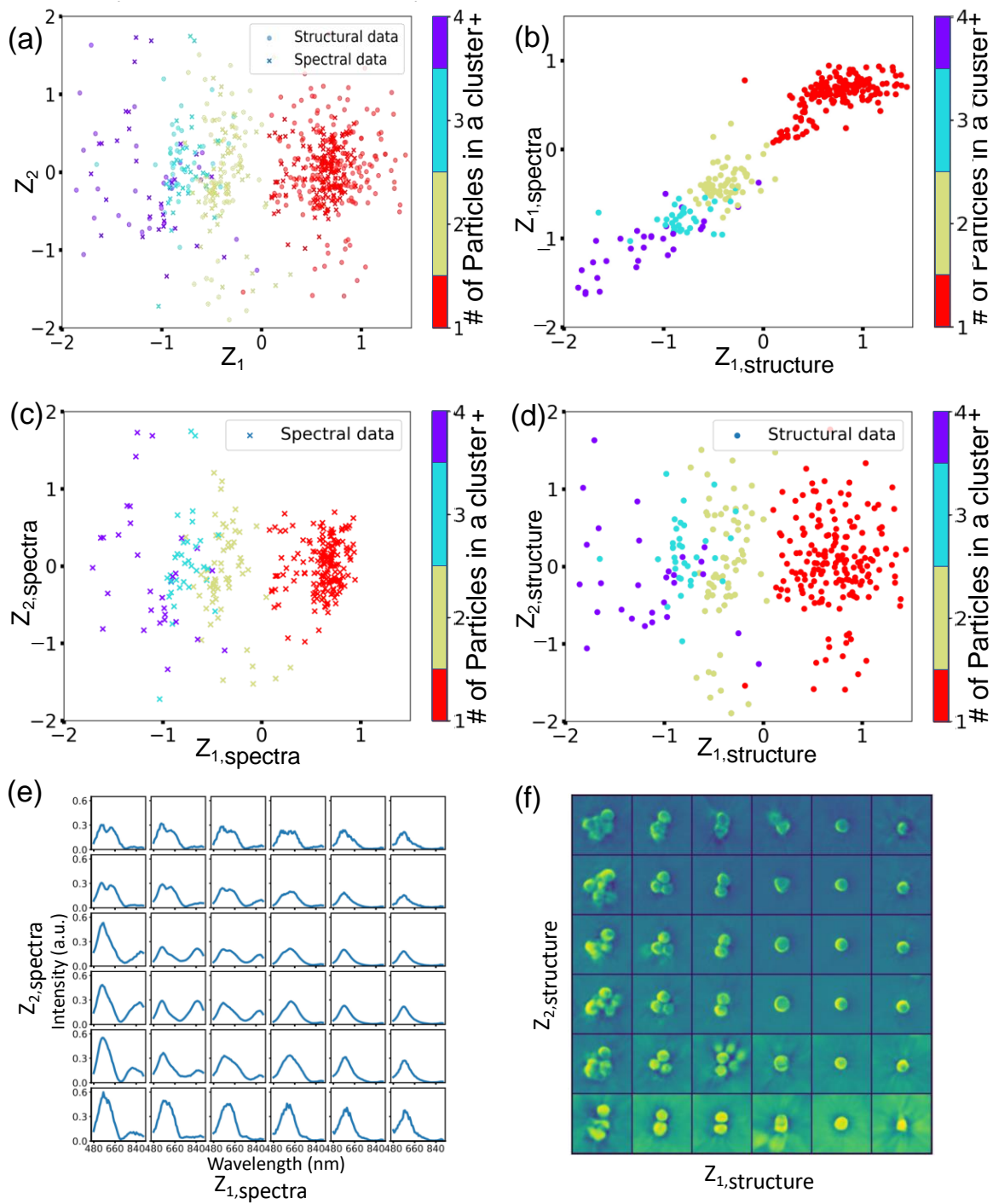

**Figure 3. Performance of the dual-VAE.** (a) Combined latent space, with star- and circle-shape points representing the spectral and structural data, respectively. (b) First latent variables of structure and spectra datasets. Individual latent space of (c) spectral and (d) structural datasets. The learned manifold representation of (e) spectral latent space and (f) structural latent space.

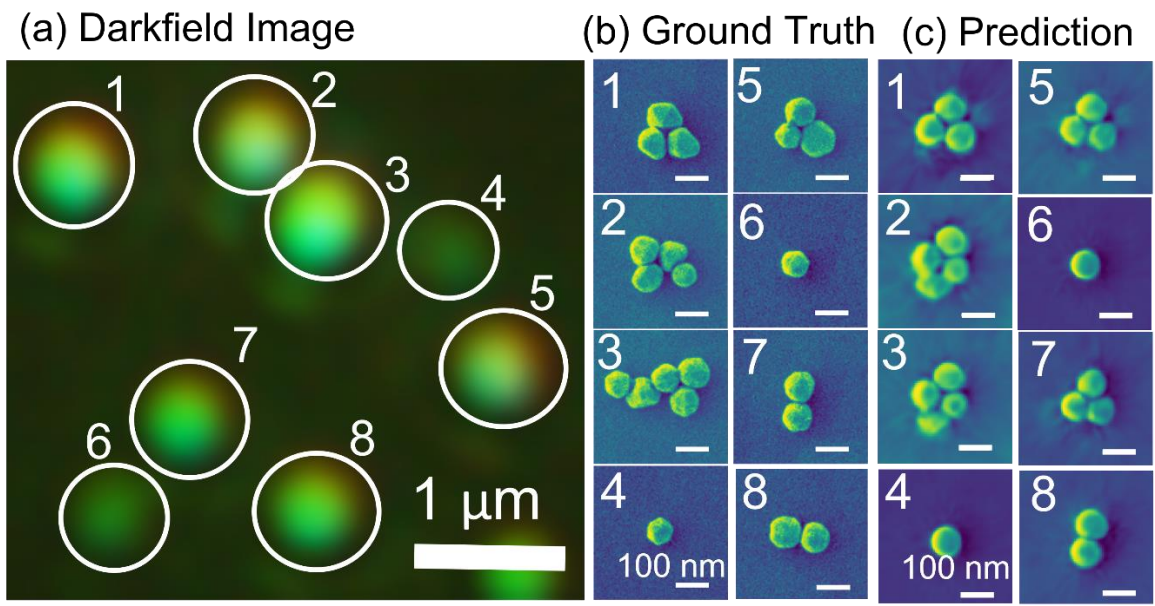

**Figure 4. Examples of the dual-VAE model's performance on an area of the test datasets**. (a) Darkfield scattering image. The (b) ground truth and (c) dual-VAE model prediction of each particle geometry based on the hyperspectral data. The scale bar is 1 μm in (a) and 100 nm in (b-c).

Fig. 3 suggests that the dual-VAE can predict the cluster geometry based on the hyperspectral data alone. To test the performance of the dual-VAE model at this task, we compare the model's predicted cluster geometry based on the spectra with the known ground truth as determined by the SEM of the actual cluster. Fig. 4 shows a comparison for a small subset of 8 clusters. Fig. 4a shows a color darkfield image of the selected region, while Fig. 4b shows the correlated SEM image for each cluster in Fig. 4a. Fig. 4c shows the predicted geometry of the cluster from the dual-VAE. Given that the dimensionality of the structure data is much larger than the spectral data, and the number of training examples is smaller than the dimensionality of the data set, this task is ill-posed task from the ML perspective. Nevertheless, the reconstructions in Fig. 4c are surprisingly accurate. Comparing the ground truth in Fig. 4b and the reconstruction in Fig. 4c, we note several points: (i) For monomers (particle 4 and 6), the prediction/decoding is nearly perfect. (ii) For dimers, the reconstruction/prediction of cluster 8 is nearly as perfect, albeit the reconstructed dimer is generated for a different orientation, reflecting the intrinsic invariances in the system. However, for cluster 7, the reconstructed image shows a trimer structure, rather than dimer, regardless of the very low intensity of the "added" particle. We attribute this behavior to the non-uniform contrast within the particles on the SEM image. (iii) The dual-VAE decoding still seems to work well for many trimers although the exact shape of each individual particle has not been preserved completely. However, the overall shape of the cluster is well maintained. (iv) The accuracy of the prediction further decreases for more complex tetramers and pentamers (cluster 2 and 3). This decrease may be ascribed to the complexity of particle geometry, the relatively small number of datasets for larger clusters ($n \geq 4$) used to train the model, and the fact that some details do not affect the photonic responses of particles significantly especially with bigger particles (i.e. the difference between a pentamer and a tetramer is less significant than a monomer to a dimer or dimer to trimer). We consider the statistical performance of this approach more rigorously after we examine a subset of the inverse problem (predicting spectra based on the image).

Although our main goal at present is reconstruction of the structural data based on the spectral data in order to determine the structure sub-diffraction-limit structures, we also used this dataset to evaluate the performance of dual-VAE model in predicting the spectral information from the structural data to determine the optical properties of the gold clusters based on their SEM images. As shown in Figure S3, we compared the spectra predicted by the dual-VAE with the ground truth from the hyperspectral image. (i) For clusters consisting of spherical particles, cluster 3, 6, 8 for example, the dual-VAE model prediction works well. (ii) The dual-VAE model can achieve better prediction on the transverse resonance peak at around 560 nm than on the longitudinal plasmon peaks, with the likely reason being that the longitudinal plasmon peak depends strongly on the distance between two particles, whereas the transverse resonance is relatively static, as we have already mentioned above. (iii) We point out that, in our case, the model still faces some challenges, as it is not easy to predict the spectra information (which, in essence, is determined by the cluster structures in three dimensions) out of SEM images that only display structure information in two dimensions. For example, cluster 1 and 2 consist of pentagonal and triangular shaped particles, while the model predicts them as spherical. The reason is that these particles overlap with other particles at the edges, thus demonstrating themselves as spherical in two-dimensional SEM images. Precise prediction of the spectra without knowing the exact 3D shape is a difficult hard task even for a human, therefore, we think that the dual-VAE model predictions are promising.

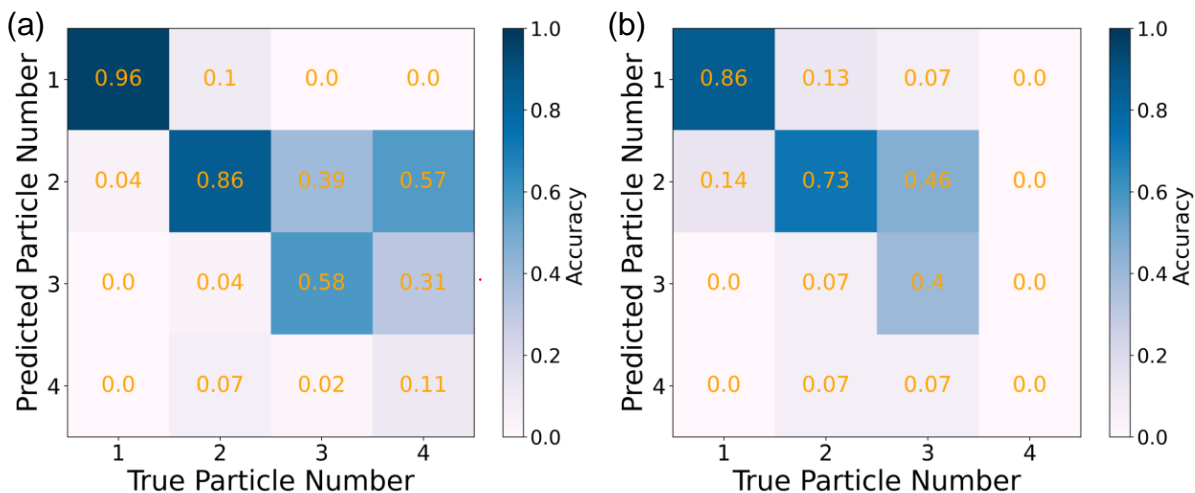

**Figure 5. Accuracy of the dual-VAE structural prediction.** Our model confusion matrix for structure prediction is based on particle numbers in a cluster in (a) our datasets, and (b) sampled literature datasets. Predicted particle numbers are our model prediction.

While Fig. 4 examines a typical region of the image to help visualize the performance of the dual-VAE, we also examined the fidelity of the reconstructions with more statistical rigor. To examine the reliability of our dual-VAE model, we quantitatively analyzed the error between the predicted structure (Figure 5) / spectra (Figure S4) against the ground truth of the clusters as determined by examining the corresponding SEM. We started from calculating the accuracy of the predicted structural data by comparing the predicted number of particles constituting each cluster with the ground truth in each cluster category. We have 651 clusters (388 monomers, 139 dimers, 89 trimers, and 35 tetramers) for testing and summarized the results in the confusion matrix shown in Figure 5. For example, out of 388 ground truth monomers, the model correctly identified 374 (374/388=0.96) objects as monomers, while the remaining 14 (14/388=0.04) were incorrectly classified as dimers. Fig. 5a shows that model achieves better prediction accuracy for monomers (0.96), and dimers (0.86), than it does for trimers (0.58), and tetramers (0.11). In the case it fails, the dual-VAE model sometimes predicts a dimer, when it should be trimer, or a trimer, when it should be tetramer.

We also tested our dual-VAE model against literature data (15 data for each class and 3 different sources [61–65]). Here, for literature data, the model prediction accuracy is 0.86 for monomers, 0.73 for dimers and 0.4 for trimers (See Figure 5b). Going beyond simply classifying particles as monomers, dimers, trimers, etc., we also use the structural similarity index (SSI)[66] to evaluate the quality of model reconstruction. The SI includes both the confusion matrix and the SSI results on images and shows that this dual-VAE model can predict the structural information with high precision.

Over, the dual-VAE represents a promising machine learning approach to build correlative relationships between the nanoparticle cluster geometries and their optical properties, with the particular approach of identifying sub-diffraction-limit particle geometries based solely on diffraction-limited far-field hyperspectral image data. Enforcing similarity between the latent representations of the structural and spectral variational autoencoders during the training appears to be a key factor in successful reconstructions based on the shared latent space. This work assumes that the encoded structure-property relationships can be approximated by a linear function, but the approach can be easily extended to more complicated dependencies.

We note that this approach of building structure-property relationships via common encoding is universal and can be applied to a broader range of materials science and physics problems in imaging and building structure-functionality relationships. The latter includes imaging/spectroscopic data emerging in the context of techniques such as (scanning) transmission electron microscopy – electron energy loss spectroscopy, the gamut of scanning probe microscopies and spectroscopies including atomic force microscopy, piezoresponse microscopy, and scanning tunneling microscopy, and chemical and optical imaging. Beyond imaging studies, this approach can be applied in molecular systems e.g., for the joint encoding of structures parametrized via graph, SELFIES, or SMILES representations and corresponding property vectors. In the case of optical systems, we anticipate that in the future it should be possible to reconstruct both particle orientations and interparticle distances within clusters, both by expanding the data collected, as well as expanding the training sets, likely with a concomitant expansion of the size of the latent space.

.

**Acknowledgements**

This work was supported (M.Y.Y., K.N.G., D.G., S.V.K.) by the US Department of Energy, Office of Science, Office of Basic Energy Sciences, as part of the Energy Frontier Research Centers program: CSSAS–The Center for the Science of Synthesis Across Scales–under Award NumberDE-SC0019288, located at University of Washington and performed (M.Z.) at Oak Ridge National Laboratory's Center for Nanophase Materials Sciences (CNMS), a U.S. Department of Energy, Office of Science User Facility. M.Y.Y. and D.S.G. acknowledge support from CNMS user facility, project number CNMS2021-B-00847. SEM imaging was conducted at the University of Washington Molecular Analysis Facility, a National Nanotechnology Coordinated Infrastructure (NNCI) site which is supported in part by the National Science Foundation, the University of Washington, the Molecular Engineering and Sciences Institute, and the Clean Energy Institute.

**Data Availability**

The detailed methodologies of dual VAE analysis on structure and spectra data set are established in Jupyter notebooks and are available from https://github.com/ziatdinovmax/dualVAE.

## References

(1) Karker, N.; Dharmalingam, G.; Carpenter, M. A. Thermal Energy Harvesting Plasmonic Based Chemical Sensors. *ACS Nano* **2014**, *8* (10), 10953–10962.

(2) Smith, J. G.; Faucheaux, J. A.; Jain, P. K. Plasmon Resonances for Solar Energy Harvesting: A Mechanistic Outlook. *Nano Today* **2015**, *10* (1), 67–80.

(3) Du, M.; Tang, G. H. Plasmonic Nanofluids Based on Gold Nanorods/Nanoellipsoids/Nanosheets for Solar Energy Harvesting. *Solar Energy* **2016**, *137*, 393–400.

(4) Hamed, M. S. G.; Ike, J. N.; Mola, G. T. Plasmonic Nano-Particles Mediated Energy Harvesting in Thin-Film Organic Solar Cells. *Journal of Physics D: Applied Physics* **2021**, *55* (1), 015102. https://doi.org/10.1088/1361-6463/ac24c8.

(5) P. Kulkarni, A.; M. Noone, K.; Munechika, K.; R. Guyer, S.; S. Ginger, D. Plasmon-Enhanced Charge Carrier Generation in Organic Photovoltaic Films Using Silver Nanoprisms. *Nano Letters* **2010**, *10* (4), 1501–1505. https://doi.org/10.1021/nl100615e.

(6) Yao, K.; Salvador, M.; Chueh, C.-C.; Xin, X.-K.; Xu, Y.-X.; deQuilettes, D. W.; Hu, T.; Chen, Y.; Ginger, D. S.; Jen, A. K.-Y. A General Route to Enhance Polymer Solar Cell Performance Using Plasmonic Nanoprisms. *Advanced Energy Materials* **2014**, *4* (9), 1400206. https://doi.org/https://doi.org/10.1002/aenm.201400206.

(7) Huang, X.; H. El-Sayed, I.; Qian, W.; A. El-Sayed, M. Cancer Cell Imaging and Photothermal Therapy in the Near-Infrared Region by Using Gold Nanorods. *J Am Chem Soc* **2006**, *128* (6), 2115–2120. https://doi.org/10.1021/ja057254a.

(8) M. Gobin, A.; Ho Lee, M.; J. Halas, N.; D. James, W.; A. Drezek, R.; L. West, J. Near-Infrared Resonant Nanoshells for Combined Optical Imaging and Photothermal Cancer Therapy. *Nano Letters* **2007**, *7* (7), 1929–1934. https://doi.org/10.1021/nl070610y.

(9) Choi, M.-R.; J. Stanton-Maxey, K.; K. Stanley, J.; S. Levin, C.; Bardhan, R.; Akin, D.; Badve, S.; Sturgis, J.; Paul Robinson, J.; Bashir, R.; J. Halas, N.; E. Clare, S. A Cellular Trojan Horse for Delivery of Therapeutic Nanoparticles into Tumors. *Nano Letters* **2007**, *7* (12), 3759–3765. https://doi.org/10.1021/nl072209h.

(10) Maier, S. A.; Kik, P. G.; Atwater, H. A.; Meltzer, S.; Harel, E.; Koel, B. E.; Requicha, A. A. G. Local Detection of Electromagnetic Energy Transport below the Diffraction Limit in Metal Nanoparticle Plasmon Waveguides. *Nature Materials* **2003**, *2* (4), 229–232. https://doi.org/10.1038/nmat852.

(11) Jiang; Bosnick, K.; Maillard, M.; Brus, L. Single Molecule Raman Spectroscopy at the Junctions of Large Ag Nanocrystals. *The Journal of Physical Chemistry B* **2003**, *107* (37), 9964–9972. https://doi.org/10.1021/jp034632u.

(12) L. Rosi, N.; A. Mirkin, C. Nanostructures in Biodiagnostics. *Chemical Reviews* **2005**, *105* (4), 1547–1562. https://doi.org/10.1021/cr030067f.

(13) Nikoobakht, B.; A. El-Sayed, M. Surface-Enhanced Raman Scattering Studies on Aggregated Gold Nanorods. *The Journal of Physical Chemistry A* **2003**, *107* (18), 3372–3378. https://doi.org/10.1021/jp026770+.

(14) Chen, J. I. L.; Durkee, H.; Traxler, B.; Ginger, D. S. Optical Detection of Protein in Complex Media with Plasmonic Nanoparticle Dimers. *Small* **2011**, *7* (14), 1993–1997. https://doi.org/https://doi.org/10.1002/smll.201100617.

(15) Maier, S. A.; Brongersma, M. L.; Kik, P. G.; Meltzer, S.; Requicha, A. A. G.; Atwater, H. A. Plasmonics—A Route to Nanoscale Optical Devices. *Advanced Materials* **2001**, *13* (19),

">
14